\documentclass[conference]{IEEEtran}
\IEEEoverridecommandlockouts
\usepackage{cite}
\usepackage{amsmath,amssymb,amsfonts}
\usepackage{algorithmic}
\usepackage{graphicx}
\usepackage{textcomp}
\usepackage{subcaption} 
\usepackage{xcolor}
\def\BibTeX{{\rm B\kern-.05em{\sc i\kern-.025em b}\kern-.08em
    T\kern-.1667em\lower.7ex\hbox{E}\kern-.125emX}}
\DeclareUnicodeCharacter{2212}{-}

\usepackage[draft=False]{hyperref} 
\hypersetup{
     colorlinks=true,
     linkcolor=blue,
     filecolor=blue,
     citecolor=black,      
     urlcolor=cyan,
     }

\begin{document}

\title{Dynamic sizing of required balancing capacities: the operational approach in France}


%
\author{\IEEEauthorblockN{Jonathan Dumas$^{1}$,  Viktor Terrier$^{1}$, Frédéric Bienvenu$^{2}$, Sébastien Finet$^{1}$, Nathalie Grisey$^{1}$}
\IEEEauthorblockA{$^{1}$RTE R\&D, Immeuble Window 7C, Place du Dôme 92073 LA DEFENSE Cedex, France\\
		\{viktor.terrier, jonathan.dumas, nathalie.grisey\}@rte-france.com \\
    $^{2}$fr.bienvenu@gmail.com}
}

\maketitle

\begin{abstract}
System operators employ operating reserves to deal with unexpected variations of demand and generation and guarantee the security of supply. However, they face new challenges to ensure this mission with the increasing share of renewable generation. This article focuses on the operational approach adopted by the French transmission system operator RTE for dynamically sizing the required margins in the context of the \textit{dynamic margin monitoring strategy}. It relies on continuous forecasts of the main drivers of the uncertainties of the system imbalance.
Four types of forecast errors, assumed to be independent, are considered in this approach: the errors in wind and photovoltaic power generation, production of conventional power units, and electricity consumption. Then, the required margin is the result of comparing the global forecast error, computed as the convolution of these independent errors, with a security of supply criterion. This study presents the results of this method implemented at RTE and used in real-time operation. \end{abstract}


\section{Introduction}

It is the role of the Transmission System Operator (TSO) to procure ancillary services from balancing service providers (BSPs) to ensure operational security \cite{code-eb-A23}. However, the increasing integration of renewable energy sources has raised the uncertainties TSOs face in the daily operation of power grids. TSOs manage remaining imbalances in the system by employing contracted and non-contracted power reserves supplied by BSPs. In particular, TSOs can activate up or downward operating reserves which are defined, under current European legislation \cite{eu-guideline}, by the System Operation Guidelines (SOGL): Frequency Containment Reserve (FCR), automatic Frequency Restoration Reserve (aFRR), manual Frequency Restoration Reserve (mFRR), and Replacement Reserve (RR).

Sizing reserves has been an increasingly challenging and relevant problem in system operations. Indeed, system operations face two types of uncertainty: i) component failures, also called contingencies; ii) forecasting errors in electrical consumption, wind, and photovoltaic (PV) productions. The sizing of operating reserves has been studied extensively in the literature.
The study \cite{papavasiliou2021overview} proposes classifying the methods into three levels of complexity: \textit{heuristic} methods, \textit{probabilistic} methods, and models based on \textit{bottom-up unit commitment} and \textit{economic dispatch} approaches.

Heuristic sizing methods estimate the volume of reserves based on simple system statistics, \textit{e.g.}, the standard deviation. For instance, the study \cite{6299425} proposes statistical methods which relate the distribution of capacity shortfall to a specific reserve requirement. However, these methods cannot adapt accurately to system conditions that vary significantly due to renewable energy supply.

Probabilistic methods estimate a probability distribution function of the capacity shortfall and set the operating reserve requirements at the quantile of the derived distribution to satisfy a target reliability. They depend mainly on the assumed imbalance drivers, \textit{i.e.}, the factors that are assumed to influence the distribution of imbalances in the system. Several types of probabilistic methods have been studied: i) \textit{parametric} methods assuming a particular distribution on the sources of uncertainty (\textit{e.g.}, Gamma or Gaussian \cite{maurer2009dimensioning,breuer2013expectation, morin2019probabilistic}, see the study \cite{de2019dynamic} that provides several references); ii) \textit{kernel density estimation} \cite{jost2015new}, and \textit{machine learning techniques} such as k-nearest-neighbours, quantile regression based on artificial networks \cite{jost2016dynamic}, and k-means.

Finally, bottom-up system modeling methods employ unit commitment and economic dispatch models to dispatch units to cope with system uncertainty in real-time \cite{de2019dynamic, 9640553, papavasiliou2021overview}. 
In this approach, outages and forecast errors are represented endogenously. 
However, these models are challenging to employ in practice due to the complexity of the underlying stochastic formulation. 

This paper focuses on probabilistic methods of the manual reserves (mFRR and RR), which, as argued in \cite{de2019dynamic}, propose a favourable balance between capturing the complexity of future power system operations and simplicity of implementation. In addition, the proposed approach is dynamic as it allows for continuously sizing the required margins based on the last forecast updates.
The contributions of this paper are three-fold.
First, it presents the differences between the two main strategies adopted by European TSOs to ensure the availability of sufficient balancing capacities required to manage the continuous balance between electricity generation and demand. 
Second, it presents the operational approach implemented and used by the French TSO (RTE) for dynamically sizing the required margins in the context of the dynamic margin monitoring strategy.
Finally, the results of this method are analyzed on operational data used by RTE to compute the required margins. They indicate a daily and seasonal pattern of the required margins and emphasize the crucial role of the probabilistic margin, which regularly exceeds the deterministic one.

Section \ref{sec:Balancing-strategies} presents the differences between the two main strategies adopted by European TSOs to procure balancing capacity. Section \ref{sec:sizing-method} describes the dynamic probabilistic method implemented by RTE, and Section \ref{sec:results} provides the results. Finally, Section \ref{sec:conclusions} presents the conclusions and future works.

\section{Balancing strategies}\label{sec:Balancing-strategies}

European TSOs use various methods to secure sufficient capacity to balance the system. They can be categorized into a fully \textit{contracted strategy} and a \textit{dynamic margin monitoring} approach.
The fully contracted strategy implies sizing the needs of balancing capacities in advance and then fully procuring them from BSPs, usually through day-ahead reserve markets. The selected BSPs are then responsible for organizing the dispatch of their assets to offer those capacities for activation close to real-time effectively. The advantage of this strategy is that the TSO does not have to continuously ensure the availability of balancing capacities. German TSOs, for example, use this approach since they estimate that they would otherwise lack balancing energy bids \cite{entsoe-balancing-report-2022}.

Another strategy is currently authorized in Europe \cite{code-eb-A23} and has been in place in France for years for sizing manual reserves. Instead of procuring its total need, the TSO can count on “the volume of non-contracted balancing energy bids expected to be available”. Indeed, producers and even consumers frequently have “naturally” available balancing capacities due to their risk management or simply the result of the energy market. For instance, a plant not dispatched at its maximal capacity can offer upward flexibility even if not contracted. The continuous declaration of these available capacities to the TSO is even mandatory in France. As a result, the existing RTE practice is to monitor “the available upward margin continuously” and the “downward available margin” of the French system and only ask to modify the dispatch when this margin is insufficient. This approach minimizes the impact on the dispatch, and the cost of procurement for the TSO since only the exact lacking capacity is procured. Moreover, it allows a much more accurate sizing of the need since it is updated continuously, considering the latest generation schedules and forecasts. In the fully contracted approach, the sizing is done at best on a day-ahead basis. However, the disadvantage of the RTE strategy lies in the much greater operational complexity.

This article focuses on the operational approach adopted by RTE for dynamically sizing the required margins in the context of the dynamic margin monitoring strategy by considering a security of supply criteria. The available margin estimation and the comparison of the reserve and margin strategies are out of this work's scope. The remainder of the paper refers to "margin" as "required margin" for clarity.

\section{Sizing method}\label{sec:sizing-method}

\subsection{General description}

Electricity balancing relies on inaccurate forecasting, scheduling, and modeling, which must be considered in the margin estimation. The study \cite{de2019dynamic} describes the system imbalance drivers classified in two main categories: i) the forecast and market errors; ii) the power plant and transmission asset forced outages.
The operational approach adopted by RTE for dynamically sizing the margin employs the statistical quantification of the main imbalance uncertainty drivers of these two categories: i) wind, PV, and consumption forecasts; ii) forced outages of conventional power plants (thermal-electric, hydro-electric, and nuclear power plants). Wind power, PV, and consumption forecast models are imperfect and use uncertain weather predictions. In addition, conventional generation units are subject to unexpected technical constraints preventing them from producing the scheduled amount of electricity. These forecasting and modeling errors must be quantified as they influence the balance of supply and demand. 
%
Other uncertainty drivers could be considered, such as reserve availability, network contingencies, \textit{e.g.}, unexpected loss of a transmission asset, and mismatches due to the market design, which can lead to deterministic frequency deviations \cite{HIRTH20151035}. However, they are currently considered less significant in the French system because they are sporadic, included in other uncertainties, overly minor in energy volume, or due to a lack of available data. 

\begin{figure}[tb]
\centerline{\includegraphics[width=90mm]{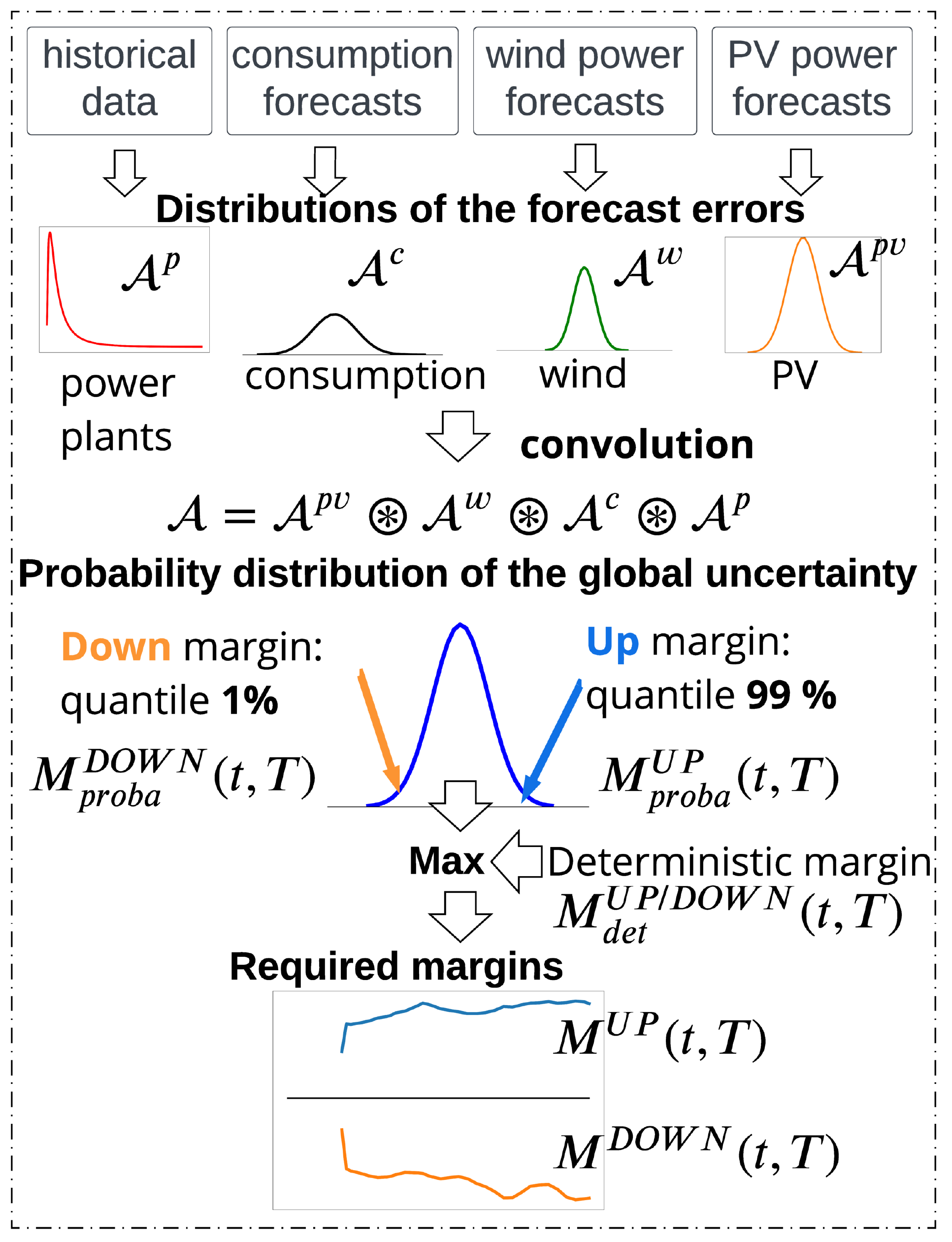}}
\caption{Illustration of the dynamic sizing method for the required margins implemented by RTE.}
\label{fig:graphical-abstract}
\end{figure}
Similarly to \cite{breuer2013expectation,de2019dynamic}, this study adopts a dynamic probabilistic approach where each uncertainty driver is dynamically estimated for each time step considered. They are assumed independent and convoluted to form a global distribution of uncertainties. Notice that margins are monitored and issued upward and downward. An upward imbalance means more production is required (or less consumption), and a downward imbalance means the opposite. Thus, upward and downward margins are computed to face the system imbalance. 
Figure \ref{fig:graphical-abstract} depicts the different steps of the computation of the margins detailed in the following sections.

\subsection{Instants definition}

\begin{figure}[tb]
\centerline{\includegraphics[width=90mm]{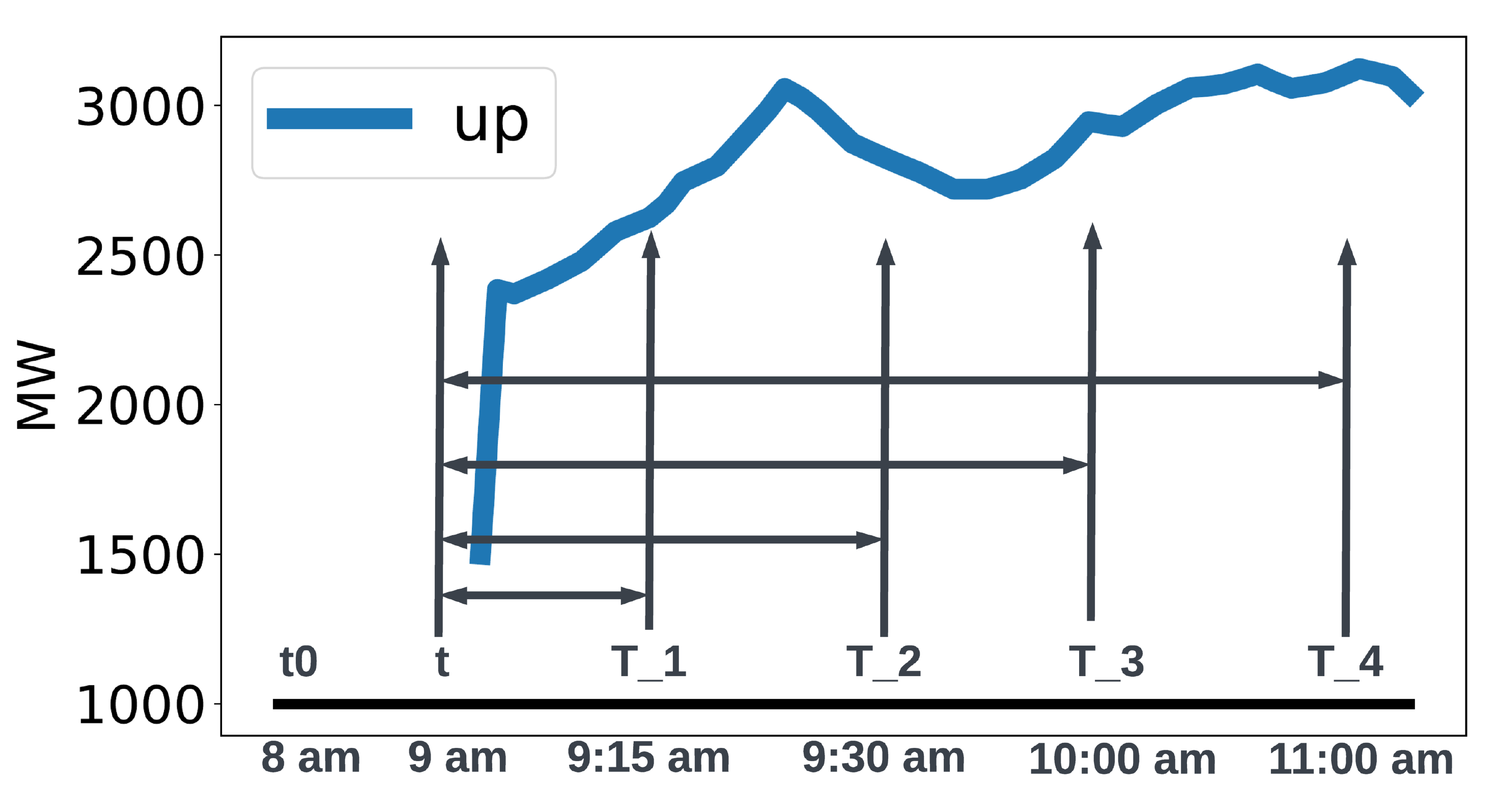}}
\caption{Illustration of the upward required margin at risk 1\%.
In this example, it is computed at $t_0 =$ 08:00 am from the instant of projection $t=$ 09:00 am to four instants of study $T=$ 09:15, 09:30, 10:00, and 11:00 am.
The black arrows point to these four specific instants of study corresponding to the anticipation periods of $T_1-t$ = 15, $T_2-t$ = 30, $T_3-t$ = 60, and $T_4-t$ = 120 minutes. 
}
\label{fig:margin-illustration}
\end{figure}

Margins are defined through three time indexes, depicted by Figure \ref{fig:margin-illustration}: 
i) T is the \textit{instant of study} where the TSO desires to assess if there are sufficiently available balancing means; 
ii) t is the \textit{instant of projection} where the operator requires to estimate the margin for T (for instance, one hour before T); 
iii) $t_{0}$ is the \textit{instant of computation} of the margin estimation, with $t_0 \leq t \leq T$.
The \textit{anticipation period} is $\Delta_T = T-t$. 
Given the information available at $t_0$, the $\Delta_T$ upward margin for T is the 99\% quantile of the global system uncertainty between T and t. RTE usually monitors margins for $\Delta_T=$ 15, 30, 60, and 120 minutes. For instance, at $t_0$ = 8 am, an operator can estimate the upward 15-minute margin required for $T=$ 9:15 am. In that case, $t=$ 9 am, and the 15-minute margin is the likely uncertainty between 9:00 am and 9:15 am. 

\subsection{Wind power and PV uncertainties}

In a nutshell, normal distributions derived from the wind power and PV quantile forecasts model the uncertainty distribution errors. Figure \ref{fig:wind-pv-power-uncertainty} in Appendix \ref{appendix:uncertainty-detail} presents a high-level description of the approaches adopted.

This approach comprises five steps.
First, the wind power and PV forecasting models implemented by RTE provide for each time step the expected value $v^{X}(t_0, T)$, the 10 \% $v_{10}^{X}(t_0, T)$ and 90 \% quantiles $v_{90}^{X}(t_0, T)$ ( $X=$ PV or wind) in the form of hourly data depending on the instant of study T, the weather forecast, and the load factor at the national level. 
Second, the PV and wind power quantile forecasts are converted to the quantile forecast errors by (\ref{eq:forecast-error}). 
Third, the PV and wind power normal forecast error distributions $\mathcal{N}^X(\mu^X,\,\sigma^X)$ are derived from the quantile forecast errors by estimating $\mu^X$ and $\sigma^X$ with (\ref{eq:mu_gen}) and (\ref{eq:sigma_gen}).
Fourth, a function $\Gamma^\text{X}$, defined by (\ref{eq:corr-wind}) for wind power and (\ref{eq:corr-pv}) for PV, learned from historical data to any instant of prediction t, allows to adjust the normal distribution $\mathcal{N}^X(\mu^X,\,\sigma^X)$ defined for $(t_0, T)$ to $(t_0, t, T)$.
Finally, the PV $q_i^{pv}$ and wind power $q_i^{w}$ quantiles $i$ \% of the forecast error is computed at a given $t_0$ for several couples $(t, T)$ with
\begin{subequations}
\label{eq:quantile-wind-pv}
\begin{align}
q_i^{w}(t_0, t,T) & = \mu^{w}(t_0,T) - \frac{\Gamma^\text{w}(\Delta_T)  \sigma^{w}(t_0,T)}{\Gamma^\text{w}(T-t_0)} \alpha_i, \label{eq:quantile-wind}\\
q_i^{pv}(t_0, t,T) & = \tilde{\mu}^{pv}(t_0,T) + \tilde{\sigma}^{pv}(t_0,T) \alpha_i, \label{eq:quantile-pv}
\end{align}
\end{subequations}
with $\alpha_i = \sqrt{2}  erf^{-1}(2 \frac{\omega_i}{100} - 1)$ and $\omega_i \in \left[ 0;100\right]$. 
Notice that in (\ref{eq:quantile-pv}), the correction function $\Gamma^\text{PV}$ does not appear. Indeed, it is directly applied in the estimation of the parameters of the PV forecast error distribution, which occurs in (\ref{eq:mu_gen}) and (\ref{eq:sigma_gen}), resulting in adjusted $\tilde{\sigma}^{pv}$ and $\tilde{\sigma}^{pv}$ parameters. The correction function $\Gamma^\text{w}$ is applied once the parameters $\sigma^{w}$ and $\sigma^{w}$ are estimated; thus, it is applied in (\ref{eq:quantile-wind}).

\subsection{Consumption uncertainties}

Several studies conducted at RTE demonstrated that a quantile regression provides the best results to model the consumption forecast error uncertainty. It uses the consumption forecast $v^{c}(t_0, T)$ and incorporates the public holidays based on operational experience because they significantly impact consumption. 
The consumption $q_i^{c}$ quantile $i$ \% of the forecast error is computed at a given $t_0$ for several couples $(t, T)$ with
\begin{align}
q_i^{c}(t_0, t,T) = & \bigg[ \beta_{i,0} + \beta_{i,1} v^{c}(t_0, T) + \beta_{i,2} f(\Delta_T) \delta_{\Delta_T \leq 180} \nonumber\\
& + (\beta_{i,3} f(\Delta_T) + \beta_{i,4})  \delta_{\Delta_T >180}  \\
& + \big[(\beta_{i,5} f(\Delta_T)  \delta_{\Delta_T \leq 600} \nonumber \\
& + \beta_{i,6} \delta_{\Delta_T>600}\big]  day(T) \bigg] \frac{\Delta_T}{f(\Delta_T)},\nonumber 
\end{align}
where $\beta_{i,k}$ are fixed by the quantile regression algorithm, $f(\Delta_T)= \max(30,\Delta_T)$, $\delta_{\Delta_T \leq X}$ is an indicator function equal to 1 if $\Delta_T < X$ and 0 otherwise, $day(T)$ is a piecewise linear function applied for each public holiday.

\subsection{Conventional generation uncertainties}

Conventional generation uncertainties include the uncertainties generated by thermal or hydropower plant operations. They represent the possibility of a plant being unable to provide the scheduled energy due to unexpected technical constraints such as start-up delay or outage. 
A Bernoulli distribution could model the uncertainties of each conventional power plant. Then, the global conventional generation uncertainty is derived from a convolution of all the Bernoulli distributions.
However, an internal technical study at RTE demonstrated better results using log-normal distributions.
Thus, two log-normal distributions are convoluted: i) for the plants with only positive generation values; ii) for all other plants (pumped-storage hydroelectric and power-to-gas power stations).

\subsection{Global uncertainty}

In the case of independent uncertainties \cite{5565529}, the density function of the global distribution of uncertainties $\mathcal{A}$ can be computed as the result of the convolution of the individual uncertainties $\mathcal{A} = \mathcal{A}^{pv} \circledast \mathcal{A}^{w} \circledast \mathcal{A}^{c} \circledast \mathcal{A}^{p}$, with $pv=$ PV, $w=$ wind, $c$ = consumption, and $p = $ conventional production. 
RTE has conducted several studies investigating the dependency of the uncertainty drivers considered. The results indicated the values of the global distribution with and without independence are mostly neglectable. Thus, we assume the four uncertainty drivers are independent.
The four uncertainty models previously described compute 201 quantiles by step of 0.5\% between quantiles 0.5\% and 99.5\%, allowing to compute a convolution of the global distribution $\mathcal{A}$ with sufficient accuracy. Then, the probabilistic margin $M_{proba}$ is defined given a security threshold presented in Section \ref{sec:security}:
\begin{subequations}
\begin{align}
M^{u}_{proba}(t,T) & = q_{99 \%}(\mathcal{A}), \\
M^{d}_{proba}(t,T) & = q_{1 \%}(\mathcal{A}),
\end{align}
\end{subequations}
with $q_{99 \%}$ and $q_{1 \%}$ the 1 \% and 99 \% quantiles of $\mathcal{A}$, and $u=$ upward and $d=$ downward.

\subsection{Deterministic margin}

The final margin $M(t,T)$ for a couple $(t,T)$ is the maximum between the probabilistic $M_{proba}(t,T)$ (described in the previous section) and the deterministic $M_{det}(t,T)$ margins
\begin{align}
M^{u/d}(t,T) & = \max [M^{x}_{det}(t,T), M^{x}_{proba}(t,T)],
\end{align}
with $x=$ u (upward) or d (downward).

The deterministic margin is defined for each anticipation period, and Table \ref{tab:deterministic-margins} provides the values for $\Delta_T=$ 15, 30, 60, and 120 minutes (Appendix \ref{appendix:det-margins} provides more details).
\begin{table}[tb]
\renewcommand{\arraystretch}{1.25}
\begin{center}
\begin{tabular}{rrr}
\hline \hline
\textbf{T} (min)&\textbf{Upward} &\textbf{Downward} \\
\hline
15 & 1500 & 500 \\
30 & 1614 & 607 \\
60 & 1842 & 821 \\
120 & 2300 & 1250 \\
\hline \hline
\end{tabular}
\caption{Deterministic required margins (MW).}
\label{tab:deterministic-margins}
\end{center}
\end{table}

\subsection{Security of supply criterion}\label{sec:security}

Currently, France encloses two security of supply criteria. 
First, a reliability target as defined in Regulation (EU) 2019/943 Article 25 \cite{code-eb-A25}, named "\textit{the 3-hour failure criterion}" set to 3 hours, including 2 hours of loss of load expectancy. It is used to size and design the power system.
 Second, a short-term criterion is named "\textit{the 1 \% criterion}". RTE currently uses it to size the required margins and aFRR to cover at least 99 \% of the French imbalances following Regulation (UE) 2017/1485 Article 157 \cite{code-eb-A157}.
The amount of upward and downward margins corresponds to the value of the 99 \% and 1 \% quantiles of the distribution resulting from the convolution. The quantiles 0.5 \% and 99.5 \% should instead be considered to fit appropriately to a 1 \% criterion. However, technical studies conducted by RTE indicated that the most extreme quantiles (close to 0 \% or 100 \%) weigh more than the central ones in the convolution. Nevertheless, they are also the most difficult to compute as more data is required. Thus, extreme quantiles are not selected for the short-term criterion. 

\section{Results}\label{sec:results}

The results of the operational approach implemented by RTE are illustrated on a real dataset\footnote{The required margins are continuously updated and available at \url{https://www.services-rte.com/}.}. 
It comprises the required upward and downward margins from mid-May 2021 to February 2023. The required margins are computed hourly for the following 15-minute periods. We restrict the analysis to the 15, 30, 60, and 120 minutes horizons.
These anticipation periods are selected due to the different types of frequency reserves (FCR, aFRR, mFRR, and RR which have activation delays of 30 seconds, 15 minutes, and 30 minutes) and the operational window (between one and two hours in France) where the TSO can activate upward or downward services.
Notice that the PV uncertainty is not considered in the results presented.

\begin{figure}[tb]
	\centering
	\begin{subfigure}{.25\textwidth}
		\centering
		\includegraphics[width=\linewidth]{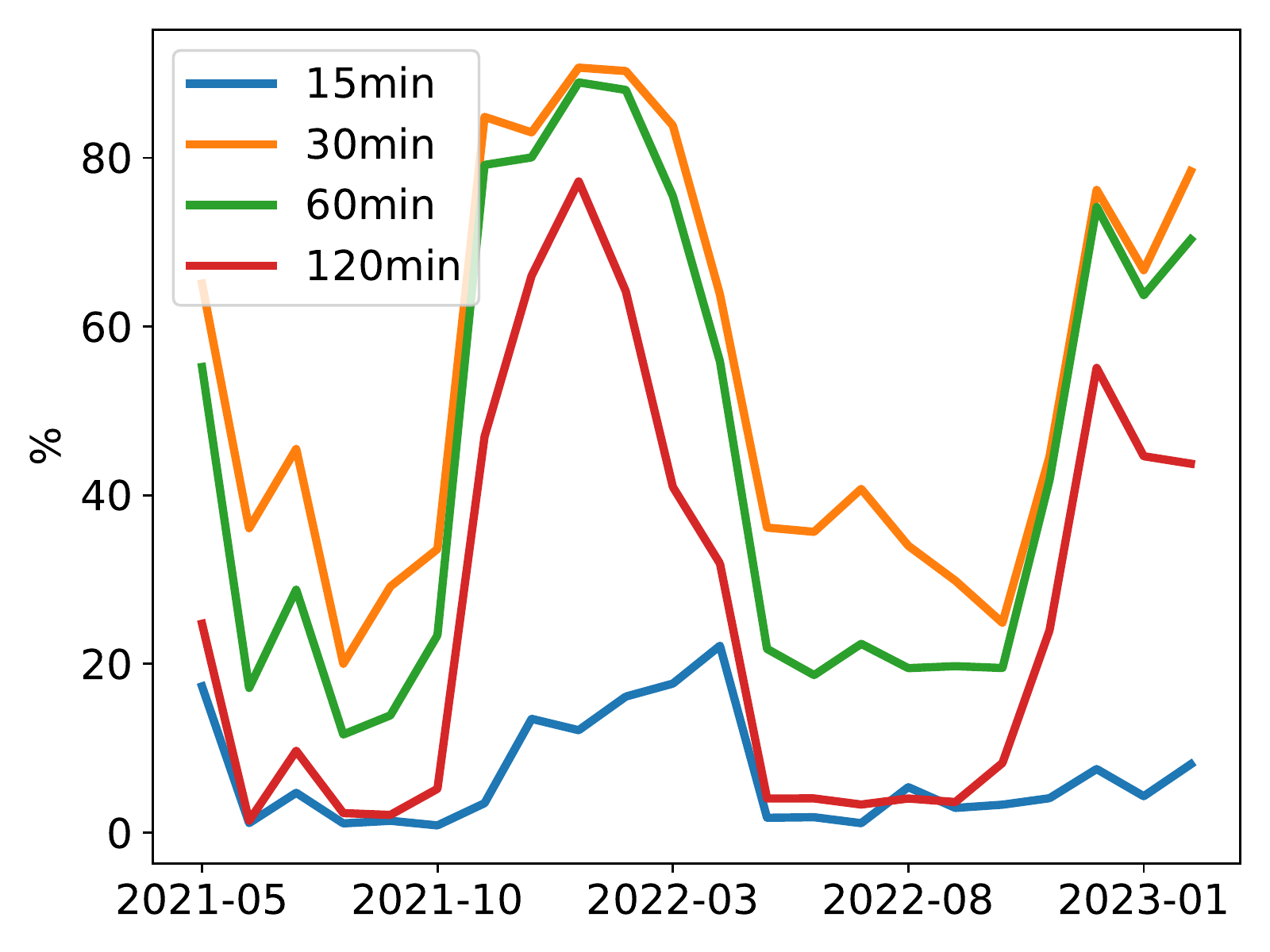}
  		\caption{Upward.}
	\end{subfigure}%
	\begin{subfigure}{.25\textwidth}
		\centering
		\includegraphics[width=\linewidth]{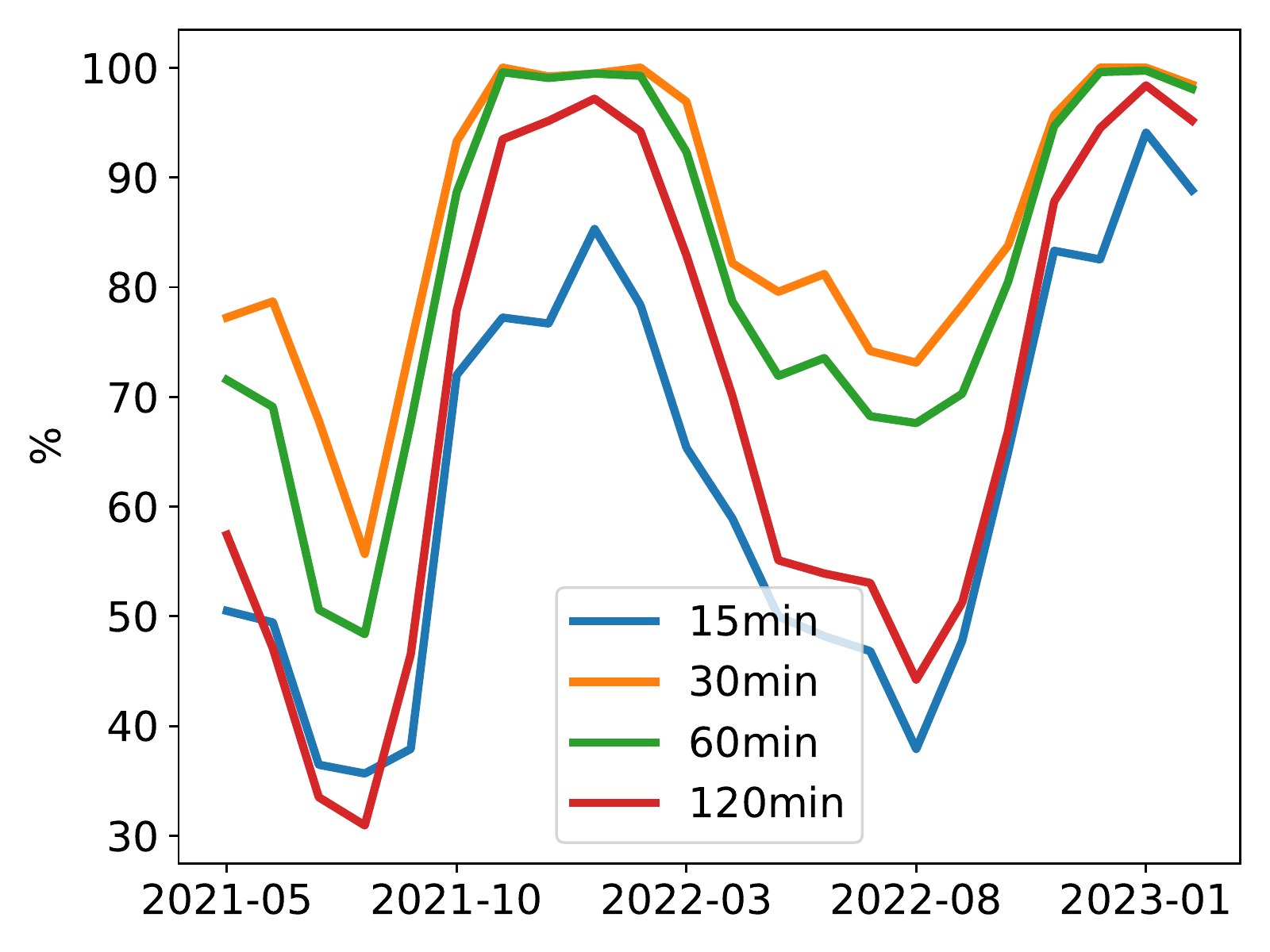}
		\caption{Downward.}
	\end{subfigure}
	\caption{Monthly ratio where the upward (left) and downward (right) required margins exceed the deterministic margin.}
	\label{fig:ratio-per-month}
\end{figure}

Figures \ref{fig:up-margin-all-data} and \ref{fig:down-margin-all-data} in Appendix \ref{appendix:results} depict the upward and downward margins over the entire dataset for these anticipation periods. 
Overall, the downward probabilistic margin exceeds slightly more frequently the deterministic margin than the upward. This is due to the lower value of the deterministic downward margin (see Table \ref{tab:deterministic-margins}). 
However, the upward and downward margins have the same trend. They both increase in winter and tend to exceed the deterministic margins as depicted by Figure \ref{fig:ratio-per-month}. 
In summer, most of the time (almost 100 \% for the 15-minute and 120-minute anticipation periods), the upward margin is equal to the deterministic threshold. However, in winter, it decreases below 30 \% except for the 15-minute horizon, which remains particularly important. Three main reasons explain this trend.
First, temperature drives mainly French electrical consumption, which leads to an average daily consumption more critical in winter than in summer and peaks of consumption in the morning and the evening. Thus, the consumption is more variable in winter than in summer, and the resulting error increases as the consumption is higher on average. Therefore, overall, it drives up the uncertainty in the consumption forecast. 
Second, additional conventional units operate in winter to supply the demand and, more occasionally, in summer. Consequently, the probability of technical failures of a power plant, such as start-up delays and unexpected shut-down, increases in winter and decreases in summer. 
Finally, similarly to electrical consumption, wind power production increases on average wind power forecast uncertainty. 

\begin{figure}[tb]
	\begin{subfigure}{.25\textwidth}
		\centering
		\includegraphics[width=\linewidth]{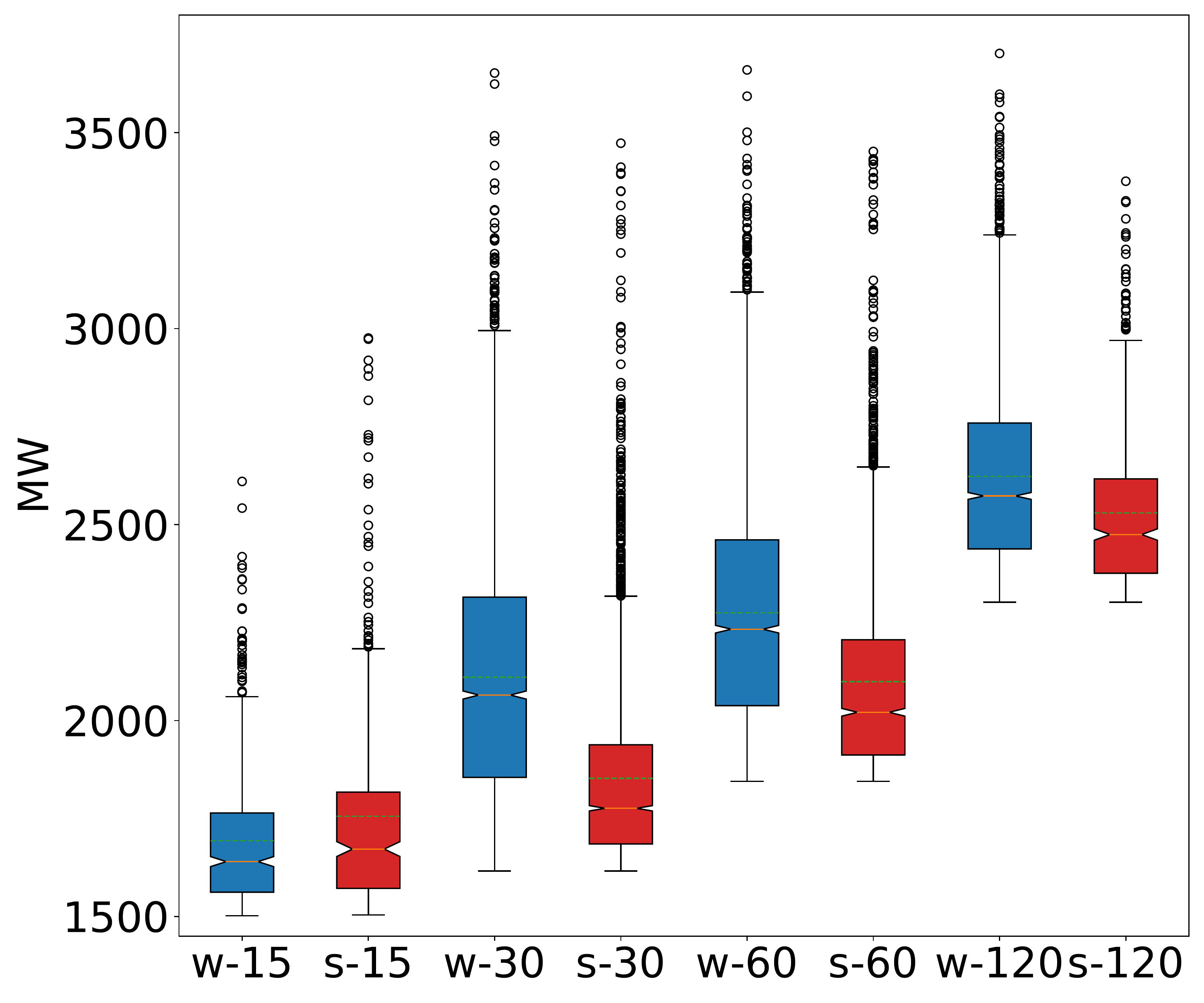}
		\caption{Upward.}
	\end{subfigure}%
 	\begin{subfigure}{.25\textwidth}
		\centering
		\includegraphics[width=\linewidth]{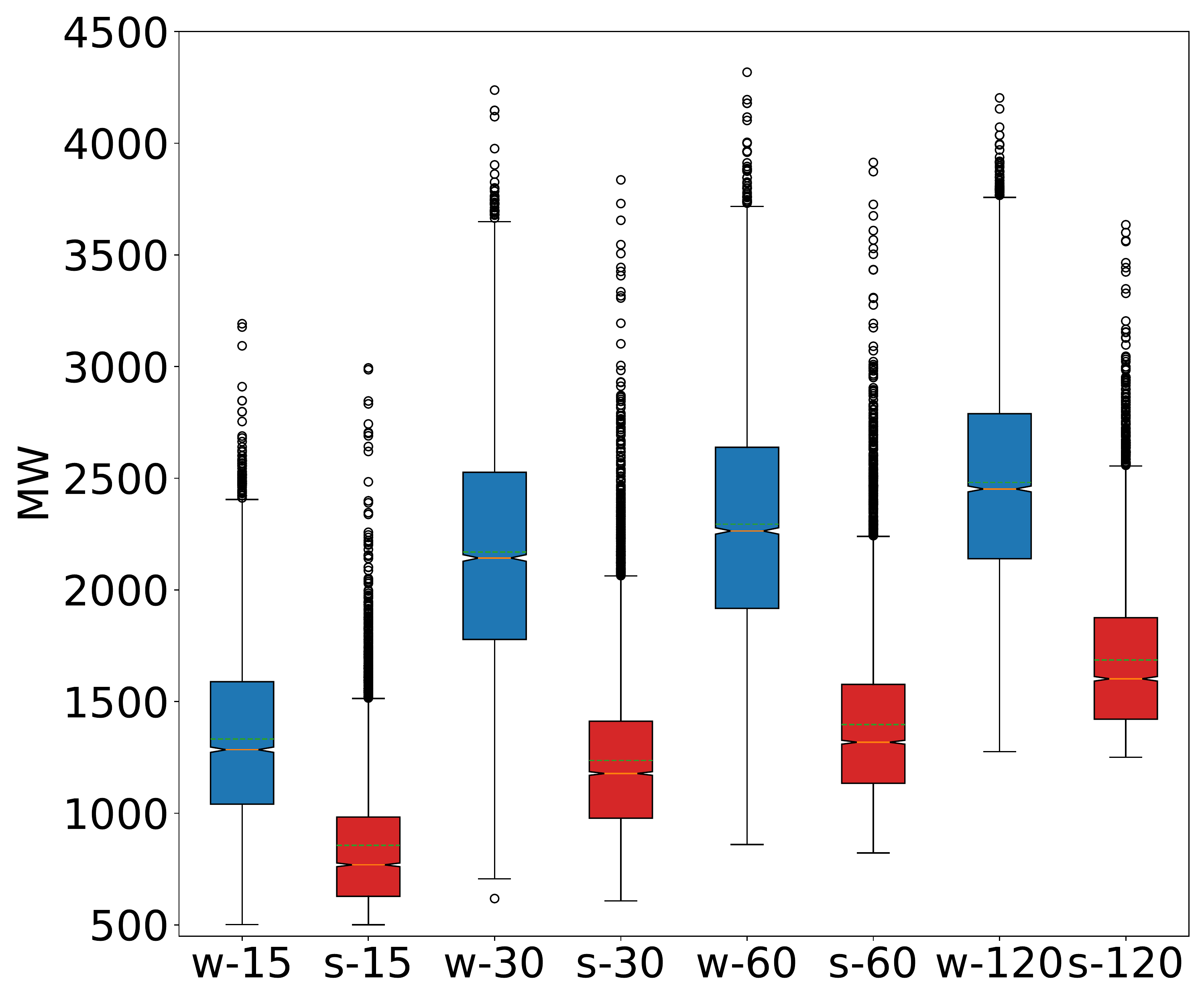}
		\caption{Downward.}
	\end{subfigure}
	\caption{Box-plots per season of the required probabilistic margin (winter in blue \textit{vs.} summer in red) for the anticipation periods 15, 30, 60, and 120 minutes. Acronyms: upward (u); downward (d); winter (w); summer (s).}
	\label{fig:boxplot}
\end{figure}

In the following, we consider only the values of the margins that exceed the deterministic thresholds. 
Figure \ref{fig:boxplot} presents statistics of the upward and downward probabilistic margins per season (winter and summer). Figure \ref{fig:boxplot-entire-dataset}, in Appendix \ref{appendix:results}, depicts the upward and downward probabilistic margins over the dataset.
Overall, the probabilistic upward margin exceeds the downward one. Furthermore, winter upward and downward probabilistic margins surpass the summer values, except for the 15-minute horizon concerning the upward margin\footnote{It is due to: i) an exceptionally high level of probabilistic upward margin in April 2022; ii) the values for the other summer months rarely exceed the deterministic margin. Thus, it increases the average over the summer months.}. 
Indeed, the wind power and consumption uncertainty distribution are identical for the upward and downward probabilistic margins. However, the conventional power plant uncertainty distribution differs between the upward and downward probabilistic margins. For the downward margin, only STEPs are considered (conventional units do not have any risk of failure to decrease power or shutdown). Overall, more conventional units are subject to outages for the probabilistic upward margin than the downward one. Thus, the resulting upward margin is higher than the downward one.

\begin{figure}[tb]
	\centering
	\begin{subfigure}{.25\textwidth}
		\centering
		\includegraphics[width=\linewidth]{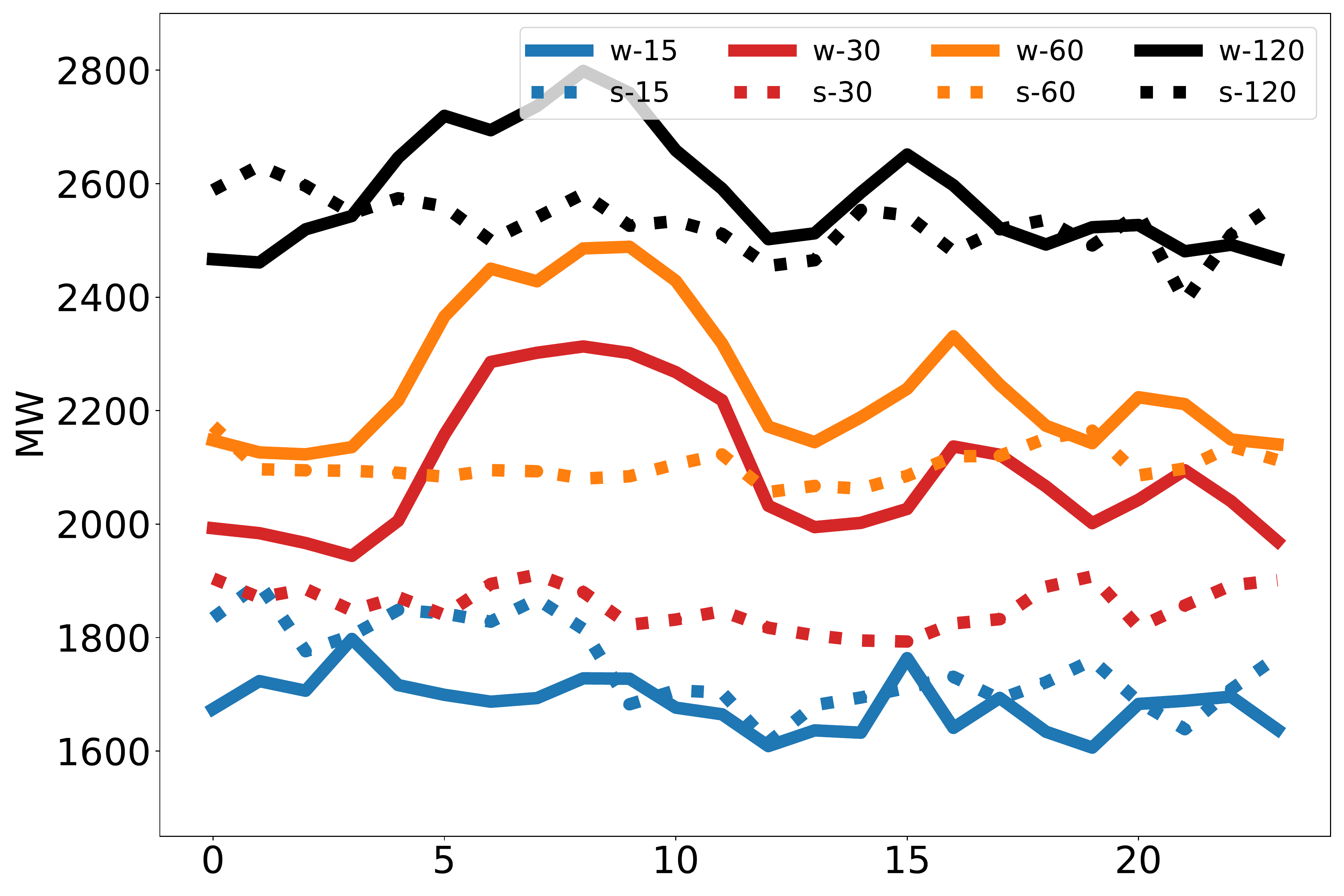}
  		\caption{Upward.}
	\end{subfigure}%
	\begin{subfigure}{.25\textwidth}
		\centering
		\includegraphics[width=\linewidth]{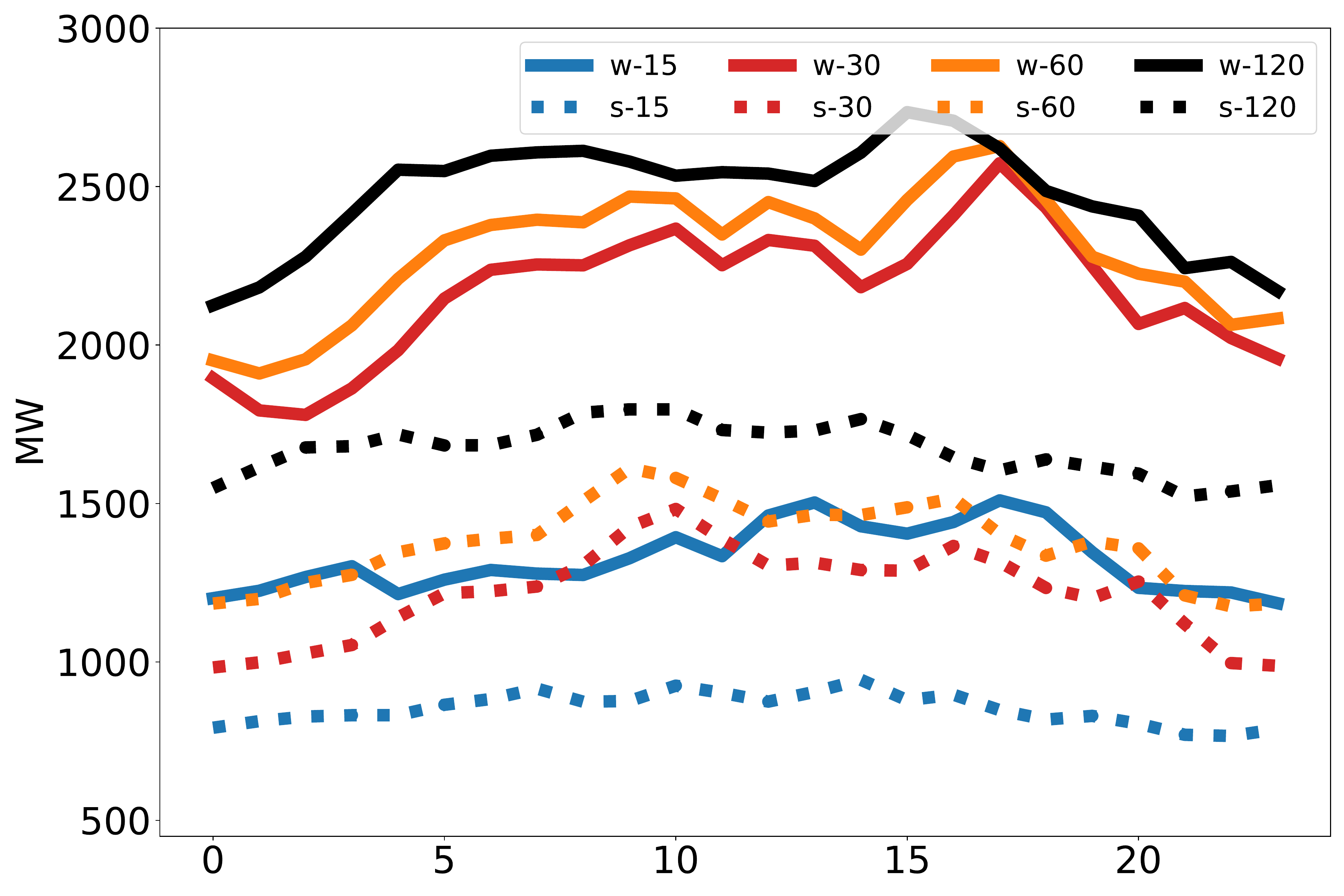}
		\caption{Downward.}
	\end{subfigure}
	\caption{Upward and downward probabilistic required margins averaged per hour and season: winter (plain line) and summer (dotted line). Acronyms: winter (w); summer (s).}
	\label{fig:value-per-hour}
\end{figure}

Finally, Figure \ref{fig:value-per-hour} depicts the upward and downward probabilistic margins averaged per hour of the day and season.
The upward and downward margins are approximately equal for all day hours in summer.
However, the upward margin in winter is more significant in the morning than during other hours. Indeed, there is a significant daily pattern with a morning consumption peak between 6 and 10 am, amplified by cold temperatures. Thus, the uncertainty of the consumption forecast increases in the morning. In addition, to balance supply and demand, more conventional power plants are started in the morning, increasing the risk of technical failure.
These results illustrate the interest in dynamically sizing the probabilistic margins to consider the uncertainty drivers optimally. In addition, it is complementary to the deterministic required margin approach, which allows taking into account specific outages or events, such as the loss of France’s most powerful nuclear power plant.

\section{Conclusion}\label{sec:conclusions}

This article presents the operational approach adopted by the French transmission system operator for dynamically sizing the required margins in the context of the \textit{dynamic margin monitoring} strategy. 
The approach considers four types of imbalance uncertainty drivers, assumed to be independent: the errors in the forecast of electricity consumption, wind and PV power generation, and the outages of conventional power units. The required margin is the result of comparing the global forecast error, computed as the convolution of these independent errors, with a security of supply criterion.
The results obtained on a real dataset indicate that the upward and downward required margins increase in winter. In particular, the upward required margin tends to be maximal during winter mornings when consumption peaks occur.
This approach demonstrates the contribution of the probabilistic required margin that allows the optimal dynamic sizing of the required margins based on the uncertainty considered. It is complementary to the deterministic approach, which allows taking into account specific outages or events, such as the loss of France's most powerful nuclear power plant or the IFA interconnectors between France and Great Britain.

Various extensions of this study could be investigated.
First, considering PV uncertainty as the installed capacity is increasing and may drive global uncertainty and the required upward margin. It could result in a more critical upward probabilistic required margin in the summer and a less frequent occurrence of the deterministic margin.
Second, investigate additional imbalance uncertainty drivers, such as deterministic frequency deviation \cite{HIRTH20151035}. They result from the way contracts are designed in liberalized electricity markets. Programs are specified as discrete step functions in intervals of 15 or 30 minutes. However, physical demand and supply changes are smooth.
Third, the risk of grid component failures increases with the growth of interconnection capacities and the development of renewable energies. This growing uncertainty driver could be considered in the near future.
Finally, the current approach sets a threshold value (1 \%) for the maximum acceptable risk. However, other decision strategies like setting another satisfactory risk level or finding a compromise between economic issues and the risk of load loss could be investigated \cite{5565529}.

\section*{Acknowledgment}

Fr\'ed\'eric Bienvenu developed the operational model. The authors would like to acknowledge the help of Godelaine De-Montmorillon and Peter Mitri in accessing the data and implementing the operational model. 


\bibliographystyle{ieeetr}
\bibliography{biblio}

\section{Appendix}\label{eem:appendix}

\subsection{PV and wind power uncertainties: mathematical details} \label{appendix:uncertainty-detail}

\begin{figure}[tb]
	\centering
\includegraphics[width=0.75\linewidth]{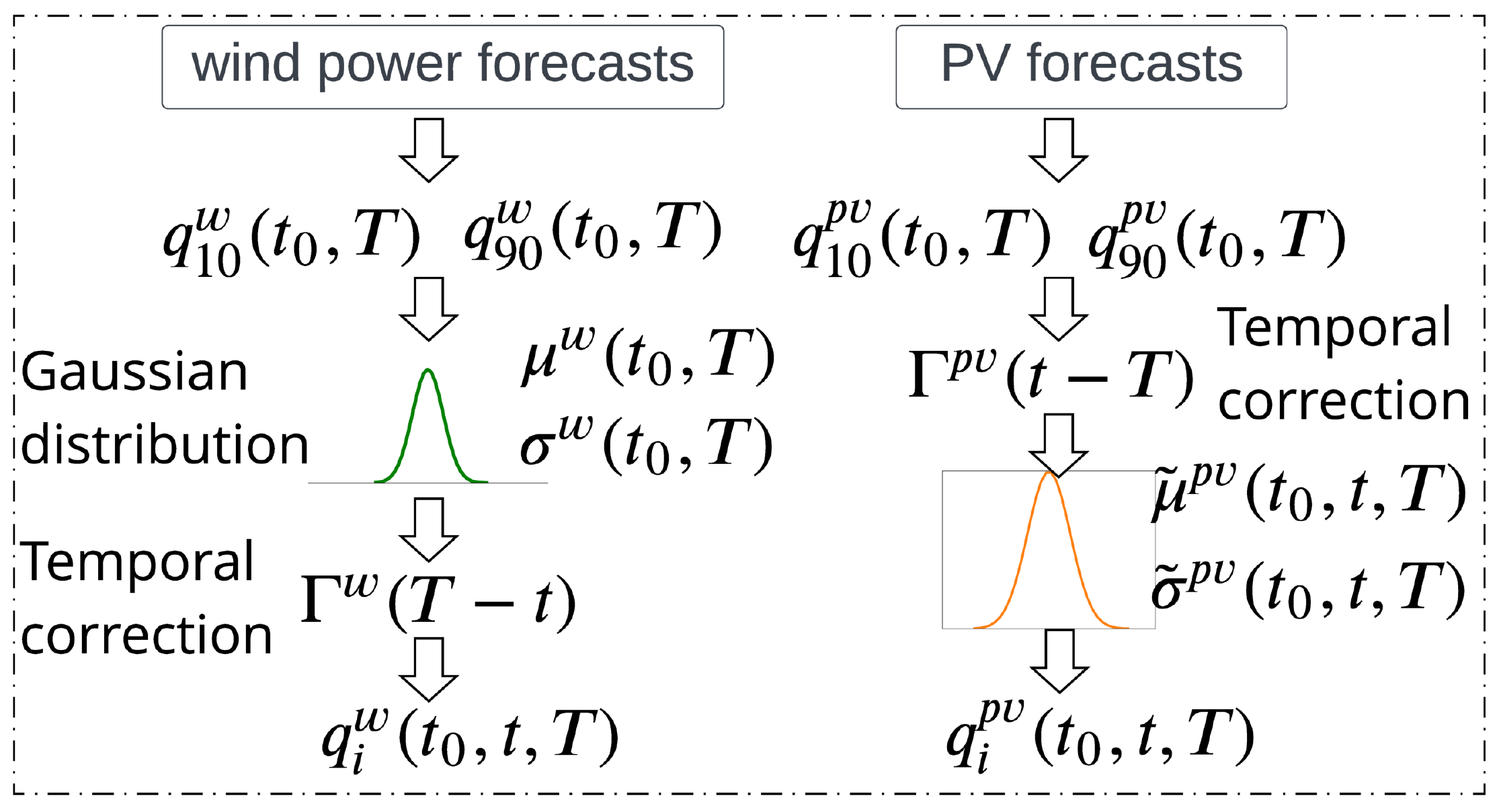}
\caption{High-level description of the wind power and PV uncertainty approaches.}
	\label{fig:wind-pv-power-uncertainty}
\end{figure}

This Appendix provides the mathematical details of Section \ref{sec:sizing-method}, and Figure \ref{fig:wind-pv-power-uncertainty} presents a high-level description of wind power and PV uncertainty approaches adopted.

The PV and wind power expected value (v) and quantile forecasts ($v_{i}$) are converted to the quantile forecast errors with
\begin{align}
\label{eq:forecast-error}
q_{100-i}^{X}(t_0,T) & = v^{X}(t_0,T) - v_{i}^{X}(t_0,T),
\end{align}
with $i=$ 10 \% and $i=$ 90 \%, and $X = $ wind or PV. Notice that, due to the subtraction, the 10$^\text{th}$ percentile of the production forecast provides the 90$^\text{th}$ percentile of the forecast error.
The PV and wind power normal forecast error distributions $\mathcal{N}^X(\mu^X,\,\sigma^X)$ are derived from these quantiles with
\begin{subequations} \label{eq:quantiles-normal}
\begin{align}
\mu^X &= \frac{q_{10}^X + q_{90}^X}{2}, \label{eq:mu_gen} \\ 
\sigma^X &= \frac{q_{90}^X - q_{10}^X}{\sqrt{2}[erf^{-1}(2 \cdot 0.9-1)-erf^{-1}(2 \cdot 0.1-1)]} .\label{eq:sigma_gen}
\end{align}
\end{subequations}
Notice that in (\ref{eq:sigma_gen}), $erf$ is the Gauss error function and $erf^{-1}$ its inverse function.

The wind power and PV error distributions $\mathcal{N}^X(\mu^X,\,\sigma^X)$, defined for $(t_0, T)$, are adjusted to any couple $(t, T)$ with a function $\Gamma^\text{X}$, learned from historical data to any instant of prediction t. For wind power, it is defined by
\begin{align} \label{eq:corr-wind}
\Gamma^\text{w}(\Delta_T) & =  \delta_{\Delta_T \leq 300} \sqrt{\frac{\Delta_T}{a^{w}}} + \delta_{\Delta_T >300} \frac{\Delta_T + b^{w}}{c^{w}}, 
\end{align}
with $\Delta_T=T-t$. $a^{w}$, $b^{w}$ and $c^{w}$ are constants set based on historical data, and $\delta_{300}(\Delta_T)$ is an indicator function equal to 1 if $\Delta_T \leq 300$ and 0 otherwise.
For PV, it is defined by
\begin{subequations}\label{eq:corr-pv}
\begin{align}
\Gamma_{10}^\text{pv}(\Delta_T) &= \left\{
    \begin{array}{ll}
          - 423 + 45\Delta_T &  \Delta_T < 2.5, \\
          −310.5-10(\Delta_T-2.5) & 2.5 \leq \Delta_T < 6, \\
          − 345.5-0.5(\Delta_T-6)& \Delta_T \geq 6,
    \end{array}
\right. \\ 
\Gamma_{90}^\text{pv}(\Delta_T) &= \left\{
    \begin{array}{ll}
          370 + 45\Delta_T &  \Delta_T < 6, \\
          640+(\Delta_T-6)& \Delta_T \geq 6 .
    \end{array}
\right. 
\end{align}
\end{subequations}
The PV correction function $\Gamma^\text{pv}$ is used to adjust the normal distribution directly in (\ref{eq:quantiles-normal}) by correcting the PV error quantiles
\begin{align}
q_{i}^{pv}(t_0,t,T) &= q_{i}^{pv}(t_0,T)\frac{\Gamma_{i}^{pv}(\Delta_T)}{\Gamma_{i}^{pv}(T-t_0)}, 
\end{align}
with $i=$ 10 \% and $i=$ 90 \%.

\subsection{Deterministic margins} \label{appendix:det-margins}

This appendix explains the values of upward and downward deterministic margins presented by Table \ref{tab:deterministic-margins}.
%
The deterministic upward margins are 1500 MW and 2300 MW for $\Delta_T=$ 15 and 120 minutes. Indeed, RTE assumes a nuclear power plant of 1500 MW is always connected to the grid. Thus, the 15-minute margin is sized for the outage of this power plant. The 120-minute margin assumes that the electrical consumption is higher than anticipated in such an outage, with a probability of 0.5. Thus, an additional margin of 750 MW is ensured to dispose of the 15-minute 1500 MW margin. Then, a linear interpolation provides the values for the other study instant.
%
The deterministic downward margin is 500 MW and 1250 MW for $\Delta_T=$ 15 and 120 minutes. The loss of the interconnection between France and the UK and an outage on pumped hydro storage stations are considered to size the 15-minute and 120 min downward margins. Then, a linear interpolation provides the values for the other study instant.
Notice that the deterministic downward margin values do not always equal those provided in Table \ref{tab:deterministic-margins}. Indeed, it depends on various parameters, such as the number of electrical dipoles in the France - UK interconnection operation. However, for the sake of clarity in this paper, we consider, at first approximation, these values.

\subsection{Results}\label{appendix:results}

This appendix presents additional Figures used to illustrate the results provided in Section \ref{sec:results}.
\begin{figure}[tb]
	\centering
	\begin{subfigure}{.25\textwidth}
		\centering
		\includegraphics[width=\linewidth]{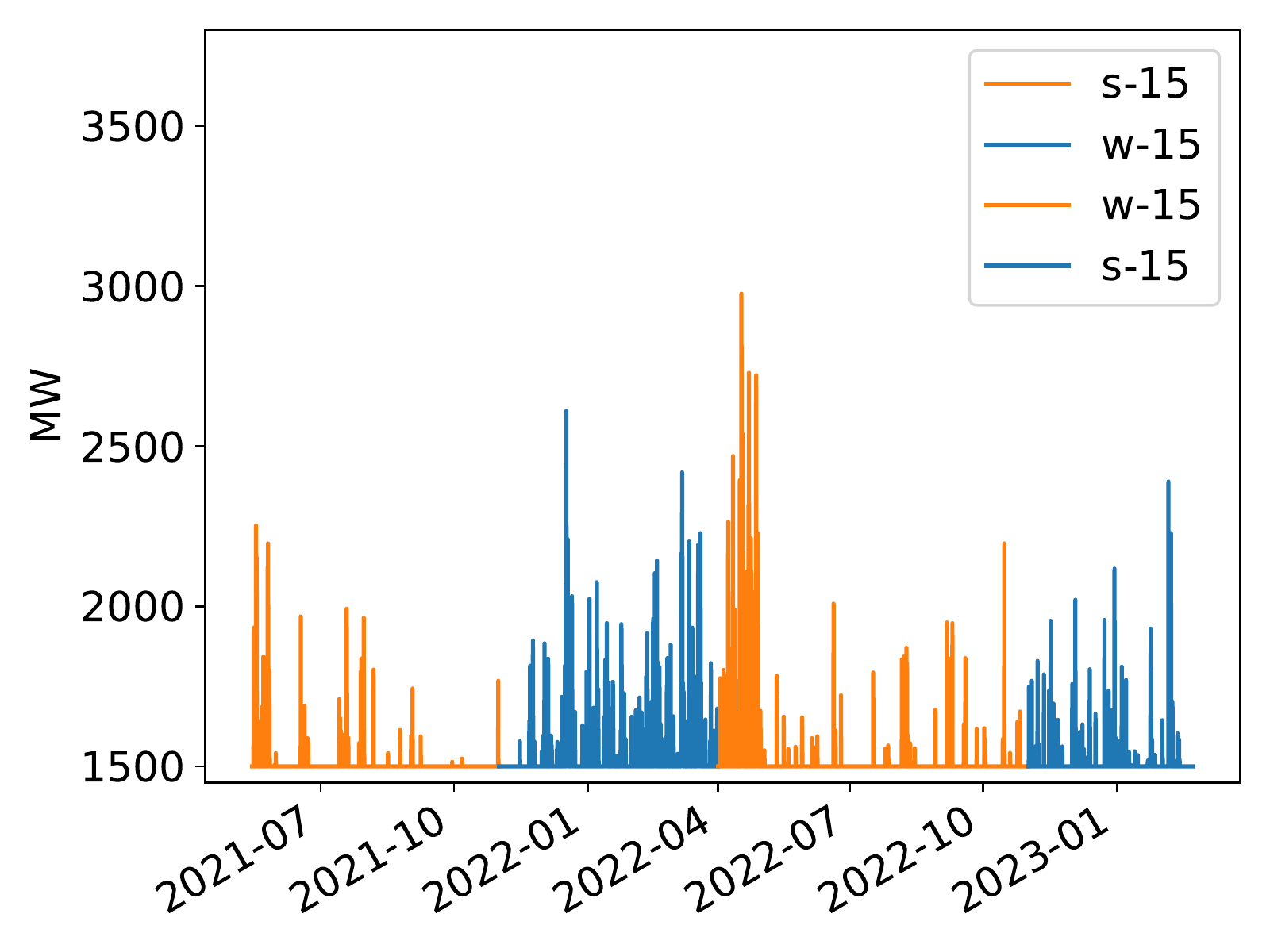}
  		\caption{15 min.}
	\end{subfigure}%
	\begin{subfigure}{.25\textwidth}
		\centering
		\includegraphics[width=\linewidth]{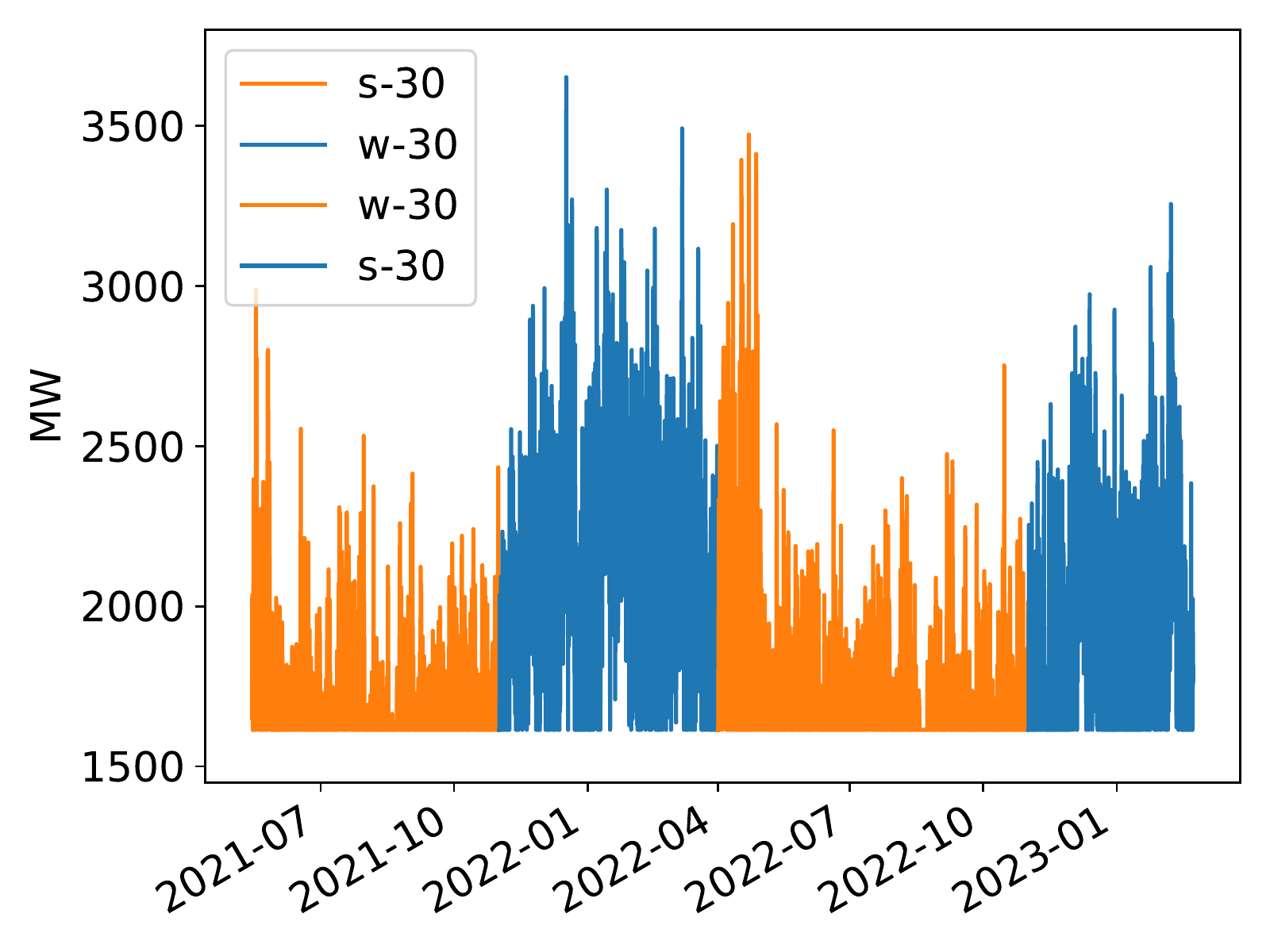}
		\caption{30 min.}
	\end{subfigure}
	\begin{subfigure}{.25\textwidth}
		\centering
		\includegraphics[width=\linewidth]{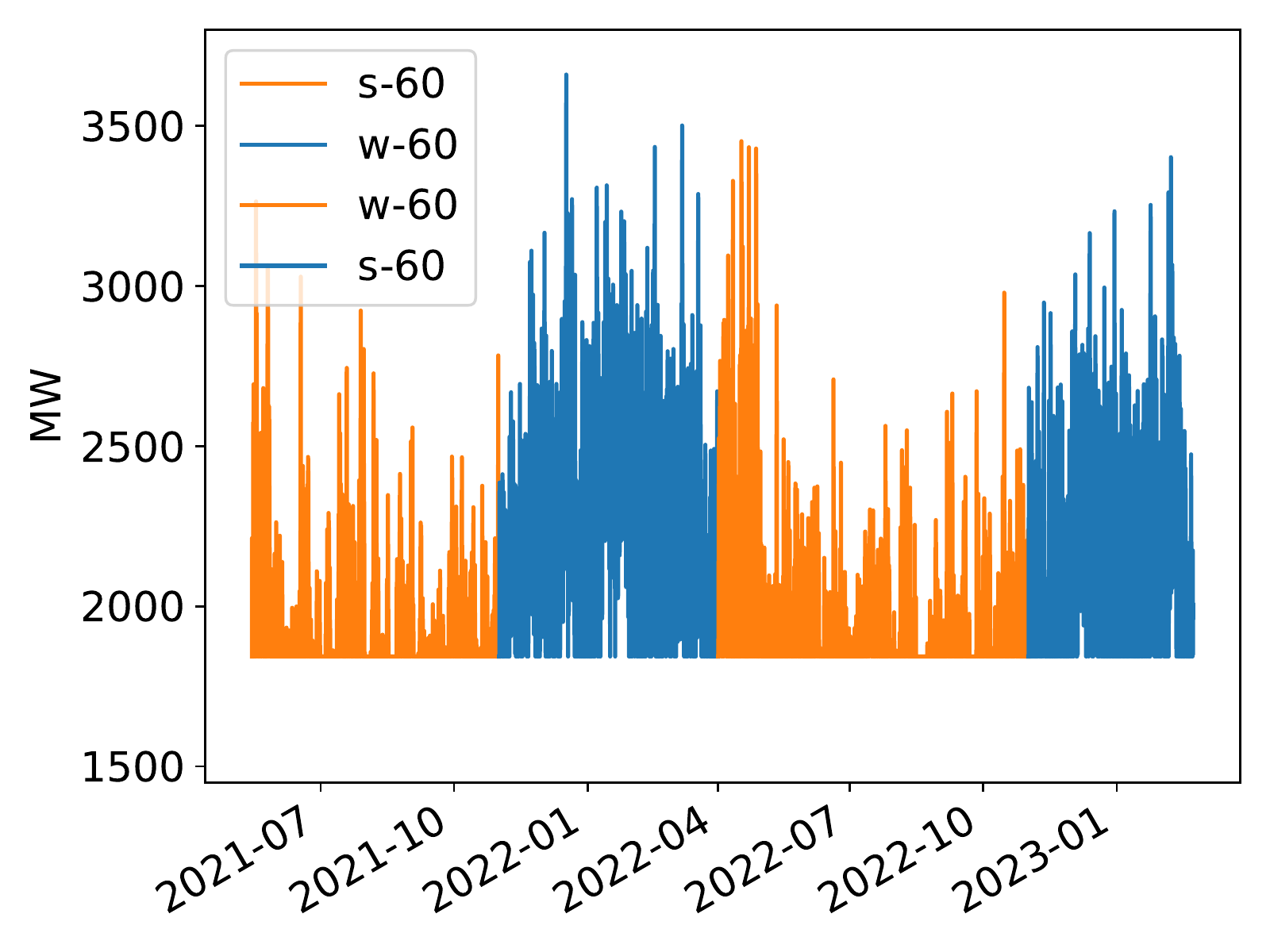}
		\caption{60 min.}
	\end{subfigure}%
	\begin{subfigure}{.25\textwidth}
		\centering
		\includegraphics[width=\linewidth]{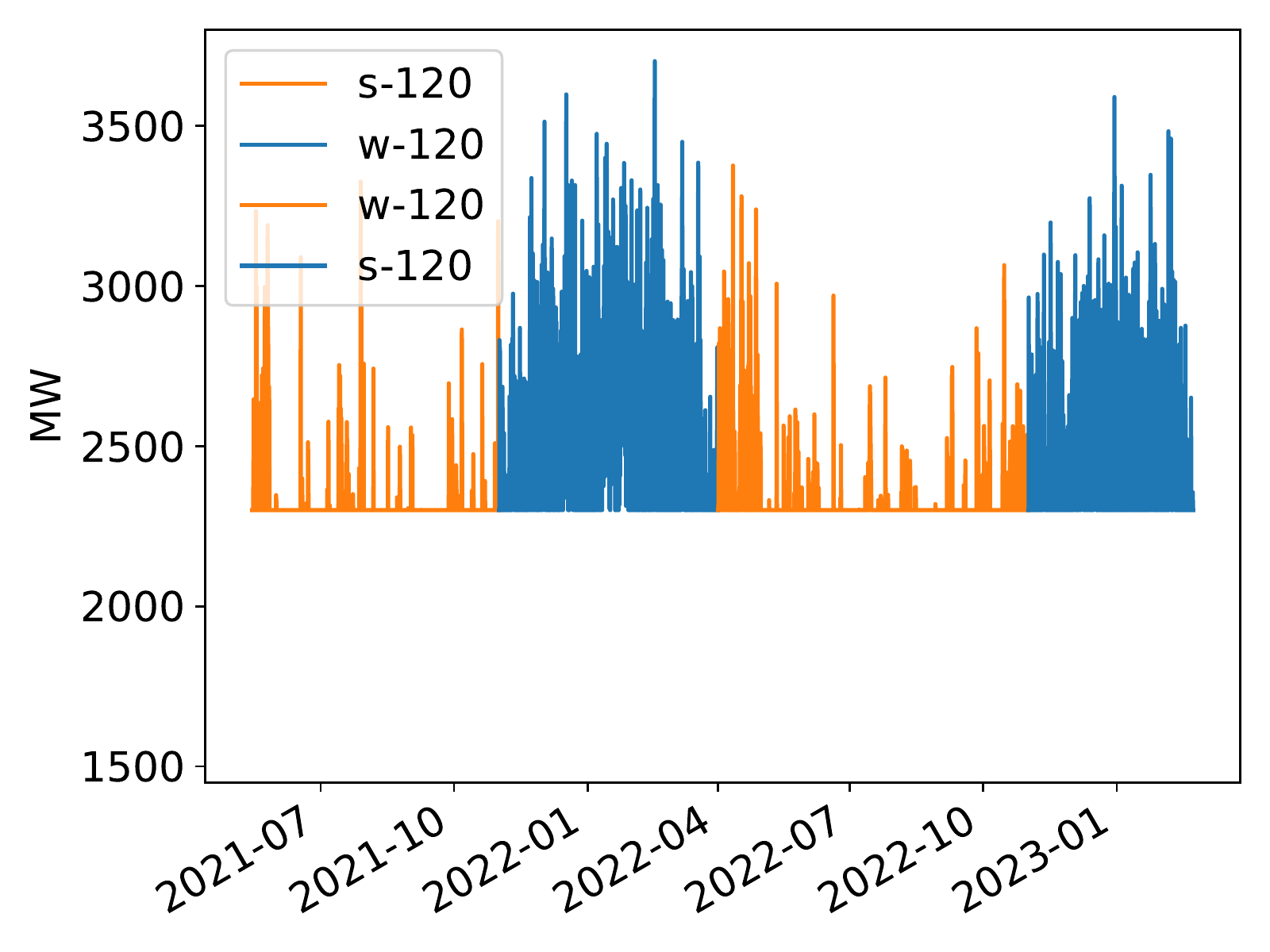}
		\caption{120 min.}
	\end{subfigure}
	\caption{Upward required margin with winter in blue and summer in orange. Acronyms: s (summer), w (winter).}
	\label{fig:up-margin-all-data}
\end{figure}
\begin{figure}[tb]
	\centering
	\begin{subfigure}{.25\textwidth}
		\centering
		\includegraphics[width=\linewidth]{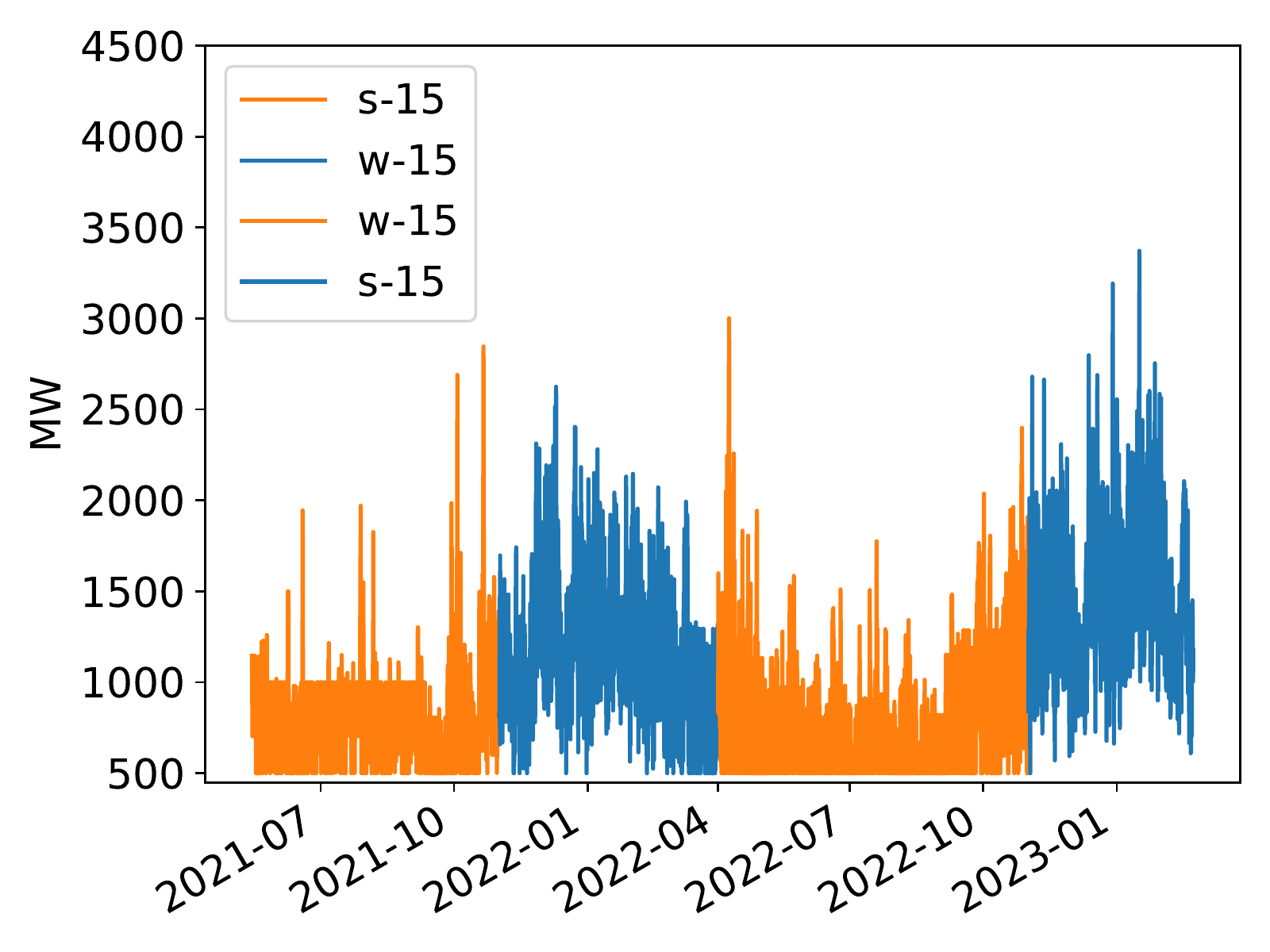}
  		\caption{15 min.}
	\end{subfigure}%
	\begin{subfigure}{.25\textwidth}
		\centering
		\includegraphics[width=\linewidth]{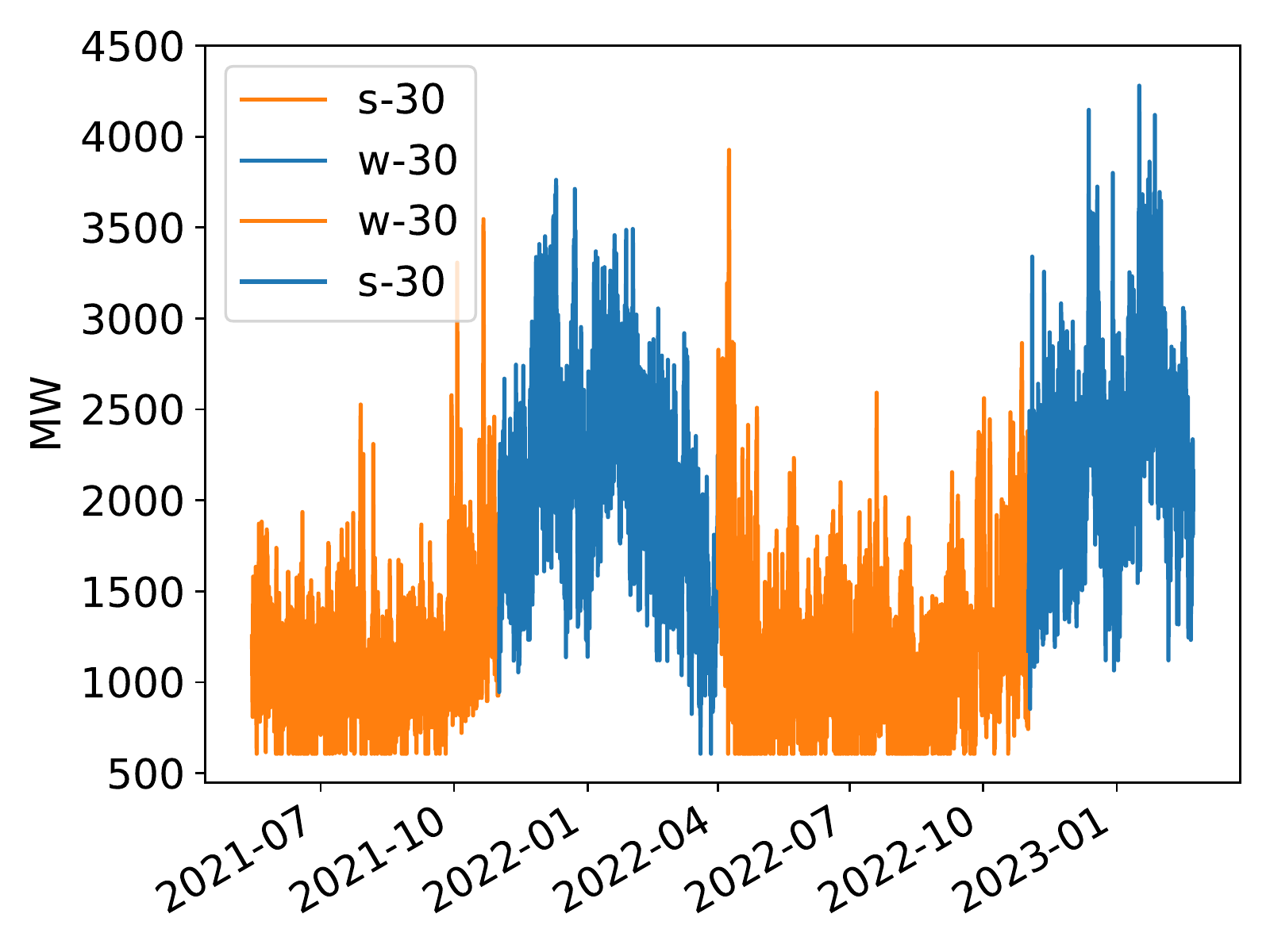}
		\caption{30 min.}
	\end{subfigure}
	\begin{subfigure}{.25\textwidth}
		\centering
		\includegraphics[width=\linewidth]{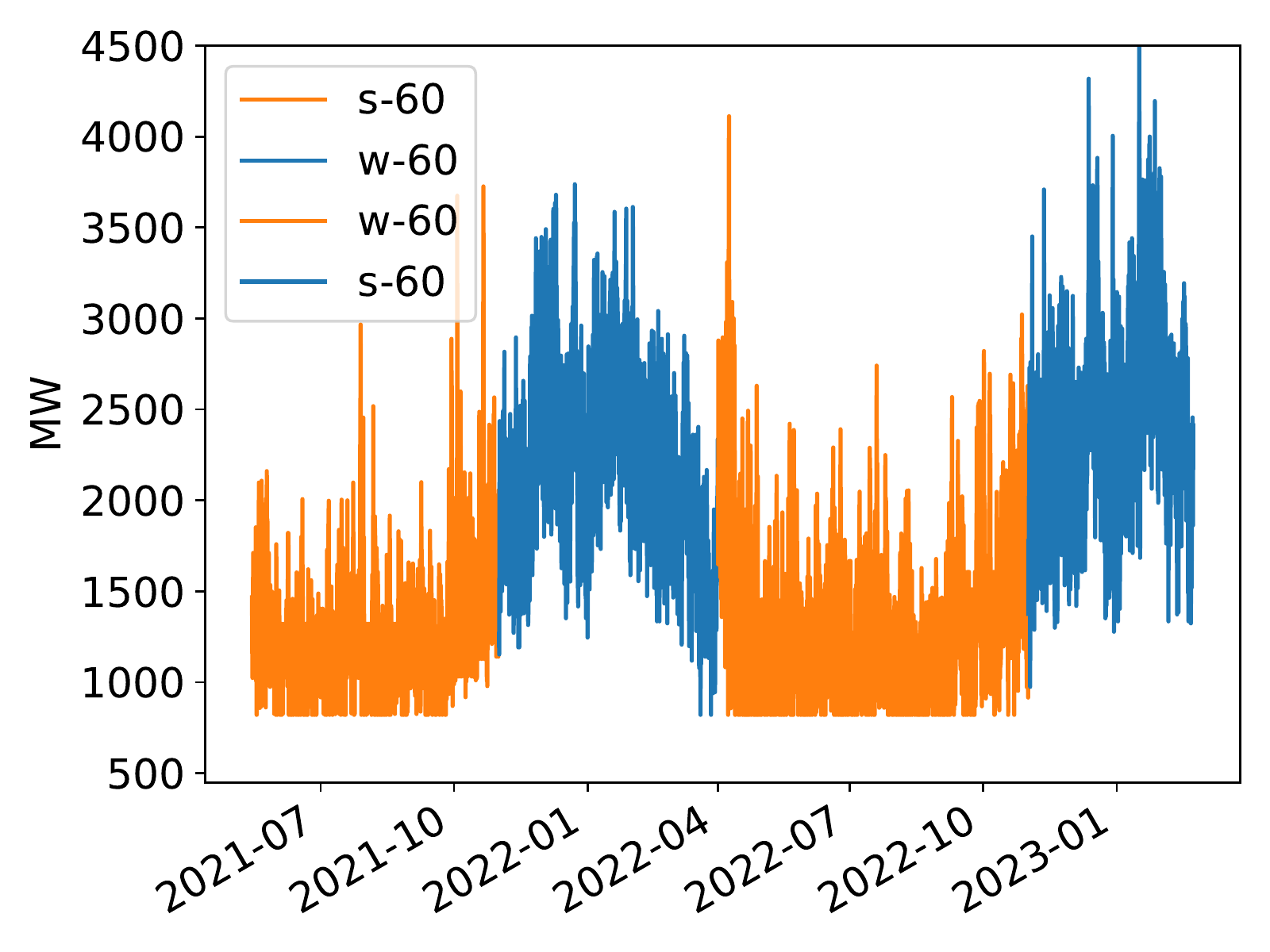}
		\caption{60 min.}
	\end{subfigure}%
	\begin{subfigure}{.25\textwidth}
		\centering
		\includegraphics[width=\linewidth]{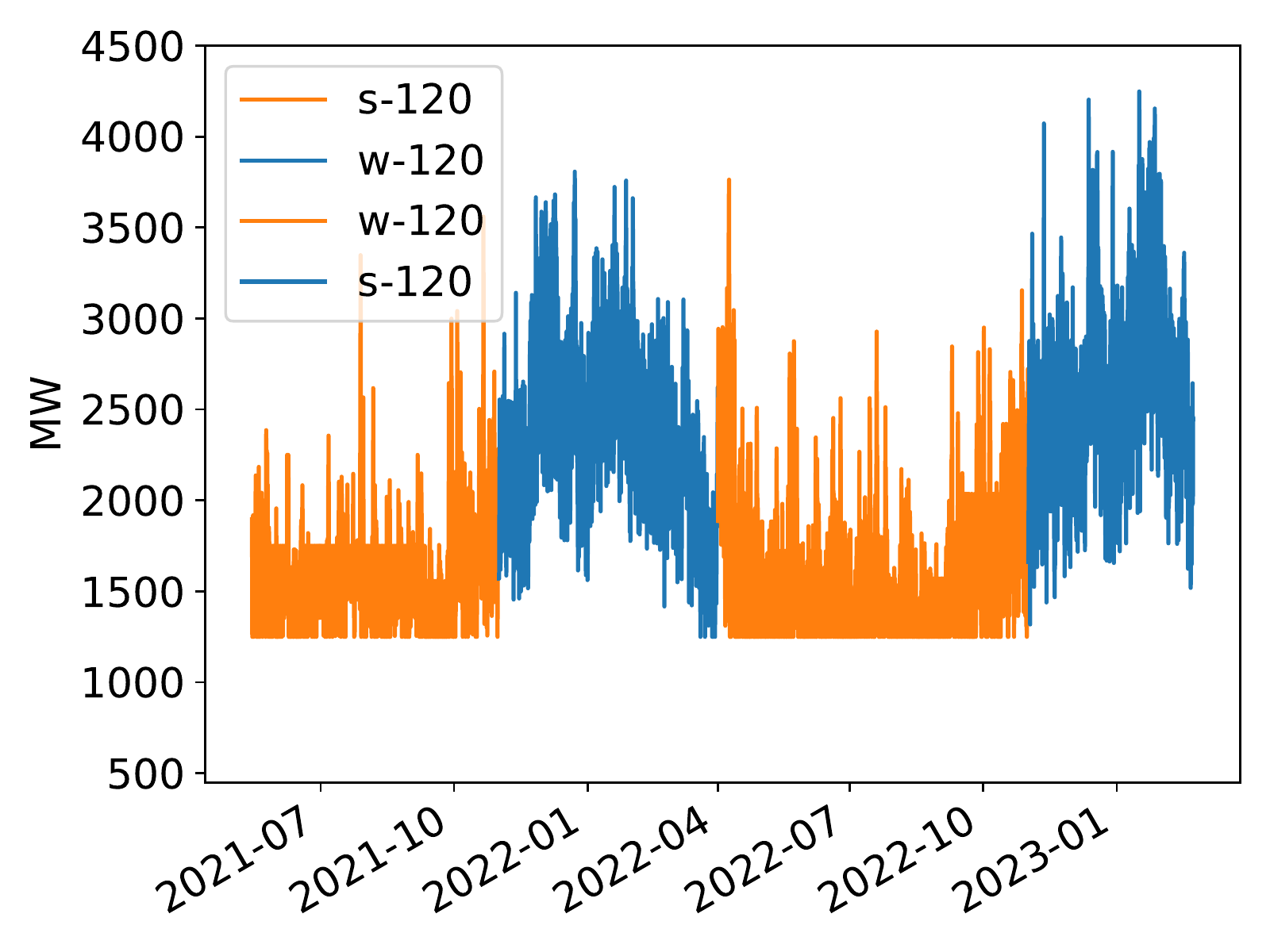}
		\caption{120 min.}
	\end{subfigure}
	\caption{Downward required margin with winter in blue and summer in orange. Acronyms: s (summer), w (winter).}
	\label{fig:down-margin-all-data}
\end{figure}
\begin{figure}[tb]
\centering
\includegraphics[width=0.75\linewidth]{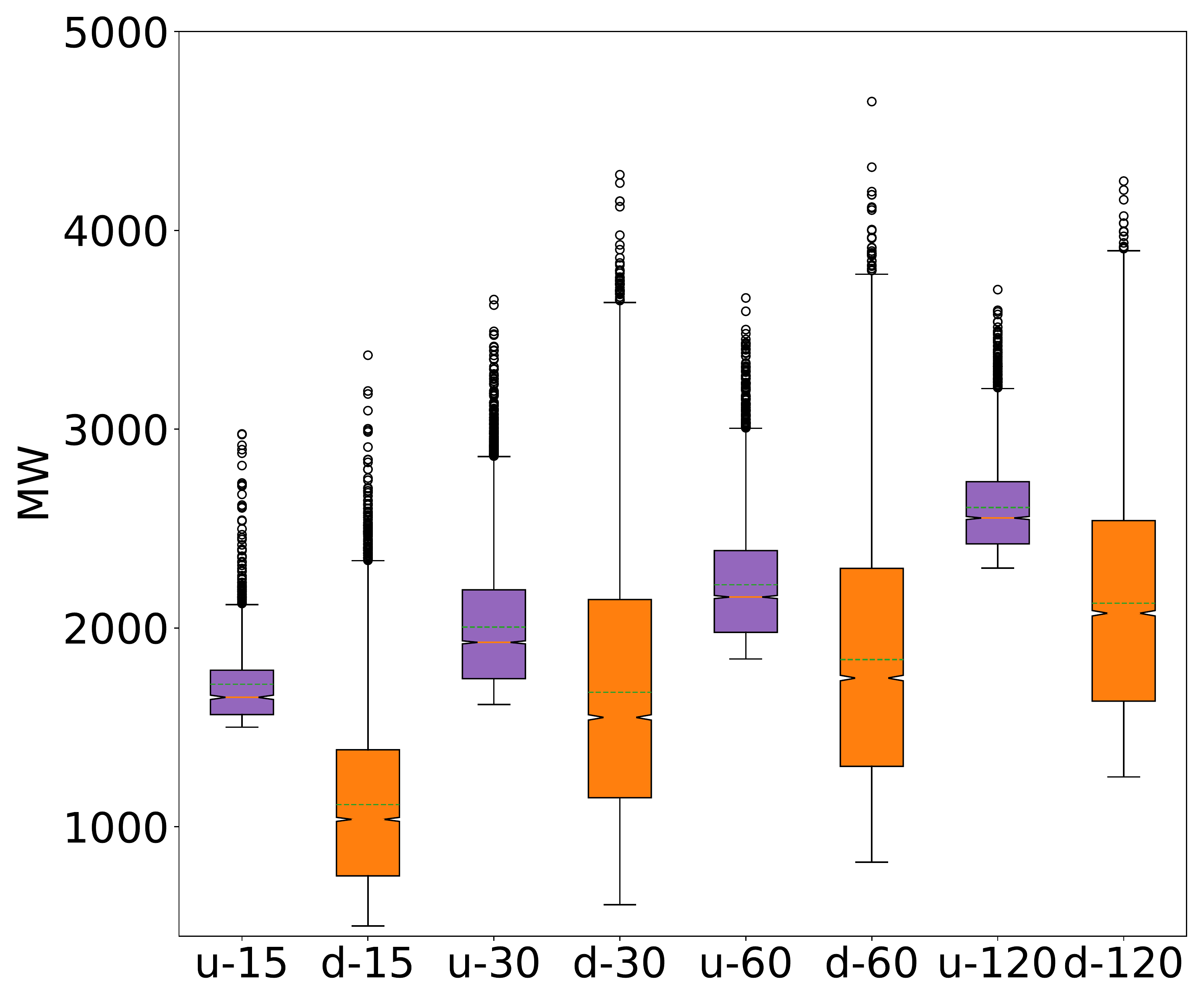}
  \caption{Box-plots of the upward (purple) and downward (orange) probabilistic required margins for the anticipation periods 15, 30, 60, and 120 minutes. Acronyms: u (upward), d (downward).} 
\label{fig:boxplot-entire-dataset}
\end{figure}

\end{document}